\documentclass[pra,10pt,amsmath,amssymb,aps,twocolumn]{revtex4-1}

\usepackage{comment}
\usepackage{epsfig}
\usepackage{graphicx}
\usepackage{dcolumn}
\usepackage{bm}
\usepackage{color}
\usepackage{ulem}
\usepackage{longtable}
\usepackage{natbib}

\begin{document}
\title{Scissors modes in generalized Gross-Pitaevskii equations}

\author{Oleksandr V. Marchukov*}
\affiliation{Institute of Photonics, Leibniz University Hannover, Nienburger St. 17, 30167 Hannover, Germany
\\
*equal contribution}
\author{Neelam Shukla*,}
\affiliation{University of Nebraska at Kearney, Kearney, NE 68849\\
*equal contribution
}
\author{Jeremy Armstrong}
\affiliation{University of Nebraska at Kearney, Kearney, NE 68849}
\author{Bastien Humbert}
\affiliation{Department of Physics and Astronomy, Aarhus University, DK-8000 Aarhus C, Denmark}
\author{Jan Arlt}
\affiliation{Department of Physics and Astronomy, Aarhus University, DK-8000 Aarhus C, Denmark}
\author{Artem G. Volosniev}
\affiliation{Department of Physics and Astronomy, Aarhus University, DK-8000 Aarhus C, Denmark}

\begin{abstract}
We investigate scissors modes in nonlinear systems with arbitrary power-law dependence of the nonlinear term. Through analytical derivation, we establish a general expression demonstrating that, in the Thomas-Fermi regime, the frequency of the scissors mode is independent of the specific form of the nonlinearity. We conclude that the scissors mode is a shear mode that does not probe the compressibility of the system, which depends on nonlinearity.
To validate our findings, we perform numerical simulations of experimentally relevant Lee-Huang-Yang (LHY) systems. Our results illustrate the transition of the scissors mode frequency from the non-interacting to the strongly interacting (Thomas-Fermi) regime. Finally, we demonstrate that the scissors mode frequency remains clearly identifiable even under strong quenches, which should facilitate the experimental observation of our findings.
\end{abstract}
\maketitle

\section{Introduction}
The nonlinear Schr{\"o}dinger equation is one of the most well-studied nonlinear equations in physics, since it describes nonlinear phenomena across a wide range of systems~\cite{Kartashov2019}. One of the primary areas where this equation is currently applied is the field of cold-atom physics~\cite{Dalfovo1999}, where it is typically called the Gross-Pitaevskii equation (GPE)~\cite{Gross1961,Pitaevskii1961}. In the cold-atom context, the GPE has been generalized in various ways~\cite{Kolomeisky2000,Thorsten2002,Pieri2003,Heiselberg2004,Choi2015,Petrov2015,Jorgensen2018,suchorowski2026}, making it interesting to compare and contrast the robustness of nonlinear physics arising from the different types of nonlinearities featured in the generalized forms of the GPE.

In this paper, we undertake such an analysis for a specific excitation mode of the degenerate Bose gas -- the scissors mode~\cite{GuryOdelin1999, Marag2000}. The name originates from the collective motion of protons against neutrons in deformed nuclei~\cite{Iudice1978, Richter1995}. In cold-atom experiments, this mode can be excited by abruptly modifying the anisotropic trapping potential~\cite{Marag2000, marago_moment_2002, Cozzini2003, Modugno2003} {\color{black}or, in the case of dipolar gases where the rotational symmetry is broken by the dipole-dipole interaction and not by the trapping potential, by modifying the direction of the external field~\cite{vanBijnen2010,FerrierBarbut2018}. The resulting oscillation frequency has been interpreted as a hallmark of superfluidity, arising from the irrotationality constraint. Indeed,} the oscillation frequency of the irrotational flow -- unlike that of the rotational mode present in the normal component -- remains finite even in the limit of a vanishing anisotropy parameter~\cite{Pitaevskii2016}. This behavior underscores its role as an indicator of superfluid properties.

In the following, we investigate scissors modes in nonlinear equations with arbitrary power-law dependence of the nonlinear term. In Sec.~\ref{sec:results}A, we derive a general analytical expression demonstrating that, in the Thomas-Fermi regime, the frequency of the mode is independent of the specific form of the nonlinearity. To illustrate our general findings, in Sec.~\ref{sec:results}B, we focus on the experimentally relevant Lee-Huang-Yang (LHY) systems~\cite{schmitt_self-bound_2016,cabrera_quantum_2018,semeghini2018,Skov2021}, which feature nonlinear interactions with a power-law dependence~\cite{Petrov2015,Jorgensen2018} that differs from the mean-field interactions typical of Bose-Einstein condensates, see Ref.~\cite{Luo2020} for a review. We explore the superfluid properties of LHY liquids by simulating the emergence of a scissors mode in a three-dimensional quantum system following a rotational quench of the anisotropic harmonic trap. We find strong finite-size effects that should be visible in current experiments that operate with  `relatively small' numbers of atoms, $N_{\mathrm{atoms}}\simeq 10^4$~\cite{Skov2021}. We also investigate the sensitivity of the scissors mode dynamics to the amplitude of the trap rotation. We find that one can identify scissors mode frequency even in strong quenches, e.g., the scissors mode dominates the dynamics even when the initial orientation of the trap is rotated by the angles $\simeq \pi/4$.  

\section{Formalism}~\label{sec:theory}
\textit{Generalized GPE.} In this paper, we consider the generalized GPE (gGPE) in three spatial dimensions 
	\begin{equation}
		-\frac{\hbar^2}{2m}\Delta\varphi + V(r)\varphi + g|\varphi|^\gamma\varphi = i\hbar\frac{\partial \varphi}{\partial t},
        \label{eq:GPE}
	\end{equation}
where $m$ is the mass of  an atom, $V$ is the external confinement; the parameters $g$ and $\gamma$ define the interaction in the system (see below). {\color{black}This equation was studied by Heiselberg~\cite{Heiselberg2004} in the context of molecular Bose-Einstein condensates and Fermi gases near Feshbach resonances. The focus of that work was on collective modes in isotropic traps and on breathing modes in deformed traps. Here, we take a step toward studying scissors modes within this gGPE.}

To study the scissors mode, the trapping potential should adopt a harmonic form in the $xy$-plane, where rotation of the trap at $t=0$ naturally excites the mode that couples with angular momentum~\cite{Pitaevskii2016}. Additionally, the harmonic trap is a natural and practical choice for cold-atom experiments.
Therefore,  for $t<0$, the external confinement is described by
\begin{equation}
V(x,y,z)=\frac{m \omega_x^2 x^2 + m\omega_y^2 y^2}{2} +V_z(z).
\label{eq:potential}
\end{equation}
This trapping potential is anisotropic $\omega_x> \omega_y$, which allows us to study rotational dynamics. The trapping potential in the $z$-direction, $V_z$, may be arbitrary~\cite{Pethick2008}.

The non-linear properties of Eq.~(\ref{eq:GPE}) are determined by the parameter $\gamma$. For $\gamma=2$, we recover the standard GPE equation~\cite{Gross1961,Pitaevskii1961}. Another prominent case is $\gamma=3$, which corresponds to the LHY liquid~\cite{Jorgensen2018}. The parameter $g$ describes the strength of the interaction. For example, for $\gamma=2$ in the context of cold Bose gases, it can be related to the boson-boson scattering length, $a$, such that $g(\gamma=2)=4\pi \hbar^2 a/m$. For $\gamma=3$ and two-component Bose gases, it can be related to an inter-species scattering length, $a_{12}$, as follows $g(\gamma=3)=128\sqrt{\pi}\hbar^2 |a_{12}|^{5/2}/(3m)$. 
The order parameter is normalized such that $\int |\varphi|^2\mathrm{d}\bm{r}=N_{\mathrm{atoms}}$.

\textit{Scissors modes.} The scissors mode is a particular solution to Eq.~(\ref{eq:GPE}) that occurs after a rapid change of the harmonic trap from the form in Eq.~(\ref{eq:potential}) to
\begin{equation}
V\overset{t=0}{\to}\frac{m \omega_x^2 x^2 + m\omega_y^2 y^2+2\theta_0 (\omega_x^2-\omega_y^2) x y}{2} +V_z(z),
\end{equation}
where the parameter $\theta_0$ defines the angle of rotation. By assumption, $\theta_0\ll 1$.
The evolution of the density of the corresponding solution resembles the rotation of a rigid body, i.e.,
\begin{equation}
|\varphi(x,y, z, t)|^2=n_0(x',y', z),
\label{eq:scissors_modes}
\end{equation}
where $n_0$ is the density of the solution of the stationary GPE at $t<0$. However, unlike the rigid body, the dynamics is irrotational (see below). The coordinates $x'$ and $y'$ are defined in the co-moving frame where the density does not appear to be disturbed
\begin{align}
x'=x\cos\theta(t)+y\sin\theta(t),  \\ y'=y\cos\theta(t)-x\sin\theta(t).
\end{align}
For scissors modes, the angle $\theta(t)$ should be a periodic function of time, which is proportional to $\theta_0$ for small perturbations.
In the limit of a very weak change of the trap, $\theta_0\to0$, it is helpful to write a Taylor expansion of the density 
\begin{equation}
|\varphi(\boldsymbol{r},t)|^2\simeq n_0(\boldsymbol{r})+\frac{i}{\hbar}\hat L_z n_0(\boldsymbol{r})\theta(t),
\label{eq:density_small}
\end{equation}
which explicitly connects the scissors dynamics to the generators of rotations, the angular momentum operator, $\hat L_z=-i\hbar (x \partial/\partial y-y \partial/\partial x)$. \\

\textit{Continuity equation.} The discussion above used the density to identify the scissors modes, motivating us to re-write the gGPE in the form of the continuity equations for a theoretical investigation of these modes. 
To this end, we insert the order parameter in the polar form $\varphi(\bm{r},t)= \sqrt{n(\bm{r},t)}e^{i\alpha(\bm{r},t)}$ into Eq.~(\ref{eq:GPE}). This leads to the equation for the density
\begin{equation}
		\frac{\partial n(\bm{r},t)}{\partial t} + \bm{\nabla}\cdot(n\bm{v}) = 0,
		\label{Continuity-equation}
\end{equation}
and for the velocity field, $\bm{v}(\bm{r},t) = \hbar\bm{\nabla}\alpha(\bm{r},t)/m$,
\begin{equation}
		m\frac{\partial \bm{v}}{\partial t} = -\bm{\nabla}\bigg(\frac{mv^2}{2}+V(\bm{r}) + gn^{\gamma/2} - \frac{\hbar^{2}}{2m}\frac{\Delta\sqrt{n}}{\sqrt{n}}\bigg)
		\label{Hydrodynamic-equation}.
\end{equation} 
If $\alpha$ is position-independent, the expression in the parenthesis is constant (chemical potential), leading to a time-independent gGPE.
Note that $\bm{\nabla}\times\bm{v}=0$, which implies that only irrotational flows can occur in the system.

\section{Results}~\label{sec:results}
\subsection{General analytical considerations}

\textit{On the existence of scissors modes.}
To demonstrate the existence of scissors modes, one needs to show that the density from Eq.~(\ref{eq:scissors_modes}), $n(\bm{r},t)=n_0(\mathbf{r}')$,  satisfies Eqs.~(\ref{Continuity-equation}) and~(\ref{Hydrodynamic-equation}) in the limit $\theta_0\to0$. To this end, we assume that $v\sim \theta_0$ and neglect all terms proportional to $\theta_0^2$. This is the so-called `linear approximation', which will be justified {\it a posteriori}.    

 In the linear approximation the equation for the evolution of the velocity field becomes
  \begin{equation}
	 	m\frac{\partial \bm{v}}{\partial t} = \bm{\nabla}\bigg( m(\omega_x^2 - \omega_y^2)(\theta(t)-\theta_0) xy \bigg).
        \label{eq:simplified_v}
	 \end{equation}
 This result is most easily derived by noticing that, by assumption, the density $n_0(\bm{r'})$ satisfies the time-independent gGPE (with $\mu$ being the chemical potential)
	 \begin{align}
	V(\bm{r'})+ gn_0(\bm{r'})^{\gamma/2} - \frac{\hbar^{2}}{2m}\frac{\Delta_{\bm r'}\sqrt{n_0(\bm{r'})}}{\sqrt{n_0(\bm{r'})}}=\mu,
    \label{eq:stationary_GPE}
	 \end{align}
     and that the Laplacian is invariant under rotations, i.e., $\Delta_{\bm{r'}}=\Delta_{\bm{r}}$.
Using the initial condition $\bm{v}(t=0)=0$, we integrate Eq.~(\ref{eq:simplified_v})  
\begin{equation}
	 	\bm{v}= (\omega_x^2 - \omega_y^2)\begin{pmatrix}
	 		y\\
	 		x\\
	 		0
	 	\end{pmatrix} \int_0^t (\theta(t')-\theta_0)\mathrm{d}t'.
	 	\label{Simplified-Hydrodynamic}
\end{equation}

Using this result and Eq.~(\ref{eq:density_small}) in Eq.~(\ref{Continuity-equation}), it is straightforward to derive the equation for the evolution of the angle, $\theta(t)$
\begin{equation}
   \frac{\mathrm{d}\theta(t)}{\mathrm{d}t}=-\Omega(x,y,z)^2\int_0^t (\theta(t')-\theta_0)\mathrm{d}t',
   \label{eq:theta_general}
\end{equation}
with the function $\Omega$ defined as follows
\begin{equation}
\Omega^2=(\omega_x^2-\omega_y^2)\frac{y n'_x+xn'_y}{y n'_x-xn'_y},
\label{eq:frequency_approximate}
\end{equation}
where $n'_x=\partial n_0/\partial x$ and $n'_y=\partial n_0/\partial y$. If $\Omega$ is constant, it defines the frequency of the scissors mode, otherwise Eq.~(\ref{eq:scissors_modes}) does not solve the time-dependent gGPE -- the scissors modes do not exist. 
Indeed, if $\Omega$ is constant, we can easily solve Eq.~(\ref{eq:theta_general}) 
\begin{equation}
\theta(t)=\theta_0(1-\cos(\Omega t)),
\end{equation}
which confirms that $\theta(t)$ and hence $\bm{v}$ are proportional to $\theta_0$ validating the derivations above, in particular, use of the linear approximation.

 If $\Omega$ is constant, we can re-write Eq.~(\ref{eq:frequency_approximate}) as follows
\begin{equation}
\frac{\Omega^2-\omega_x^2+\omega_y^2}{x}\frac{\partial n_0}{\partial x}-\frac{\omega_x^2-\omega_y^2+\Omega^2}{y}\frac{\partial n_0}{\partial y}=0.
\end{equation}
Using the method of characteristics, we conclude that this equation can be solved if and only if the density depends on the coordinates in the following way 
\begin{equation}
n_0(x,y,z)=f\left(\frac{\omega_x^2-\omega_y^2+\Omega^2}{\Omega^2-\omega_x^2+\omega_y^2} x^2+y^2,z\right),
\label{eq:density_necessary}
\end{equation}
where $f$ is an arbitrary square-integrable continuous function.
This functional form of the density is both necessary and sufficient for the existence of scissors modes. While it does not represent the most general solution to the generalized Gross-Pitaevskii equation, it is realized in two important limiting cases: very strong interactions ($g \to \infty$) and very weak interactions $(g \to 0$).
Furthermore, our numerical analysis demonstrates that this form also accurately describes systems with intermediate interaction strengths for $\omega_x\simeq\omega_y$, thereby allowing for the existence of a well-defined scissors mode across the full range of interaction parameters.

\textit{Scissors modes in a non-interacting limit}. In a non-interacting limit, the density reads as follows 
\begin{equation}
n_0(g=0)=\frac{m}{\pi\hbar}\sqrt{\omega_x\omega_y}e^{-\frac{m\omega_x x^2}{\hbar}-\frac{m\omega_y y^2}{\hbar}}.
\end{equation}
Using this density in Eq.~(\ref{eq:frequency_approximate}), we derive
\begin{equation}
\Omega(g=0)=\omega_x+\omega_y.
\end{equation}
We observe that a single scissors mode emerges in a non-interacting gas. Consequently, at zero temperature, the presence of a unique oscillation frequency upon modification of the trap serves as an indicator of Bose-Einstein condensation -- not superfluidity. In contrast, at finite temperatures, the dynamics of a thermal ensemble of particles features the frequency associated with rotational dynamics, $\omega_x -\omega_y$, which vanishes in the limit $\omega_x\to\omega_y$~\cite{GuryOdelin1999}. The appearance of this rotational mode was attributed to the breakdown of superfluidity in interacting gases~\cite{Marag2000}.

\textit{Scissors modes in a Thomas-Fermi regime.} In the strongly-interacting (Thomas-Fermi regime) the density is obtained from Eq.~(\ref{eq:stationary_GPE}) by disregarding the quantum pressure term
\begin{equation}
n_0(g\to\infty)=\left(\frac{\mu-V}{g}\right)^{2/\gamma}. 
\end{equation}
Using this result, we derive 
\begin{equation}
\Omega=\sqrt{\omega_x^2+\omega_y^2}.
\label{eq:frequency_strong}
\end{equation}
It might seem surprising that this result is independent of~$\gamma$. This result follows from the general condition that requires a specific coordinate dependence of the density, which is independent of $\gamma$. In other words {\color{black}(unlike the breathing mode~\cite{Heiselberg2004})}, the scissors mode in a harmonic trap is a pure shear mode with no density compression in a linear approximation.
Because there is no density compression, the scissors mode does not probe the internal state of the system, which depends on the specific non-linearity. 

\textit{On the average value of $\langle xy \rangle$}. 
To verify experimentally the derivations above one can measure $\langle xy\rangle$~\cite{Rossi2016}, which for the scissors modes is connected to the angle $\theta(t)$ as (cf.~Eq.~(\ref{eq:density_small}))
 \begin{equation}
	 	\langle xy\rangle \equiv \int \mathrm{d} \bm{r}  \frac{xy \; n(\bm{r},t)}{N}= \frac{\theta(t)}{N}\int \mathrm{d}\bm{r}~n_0(\bm{r})\big(x^2 - y^2\big).
	 	\label{Proportionality}
	 \end{equation}

\begin{figure}
\centering
\includegraphics[width=\linewidth]{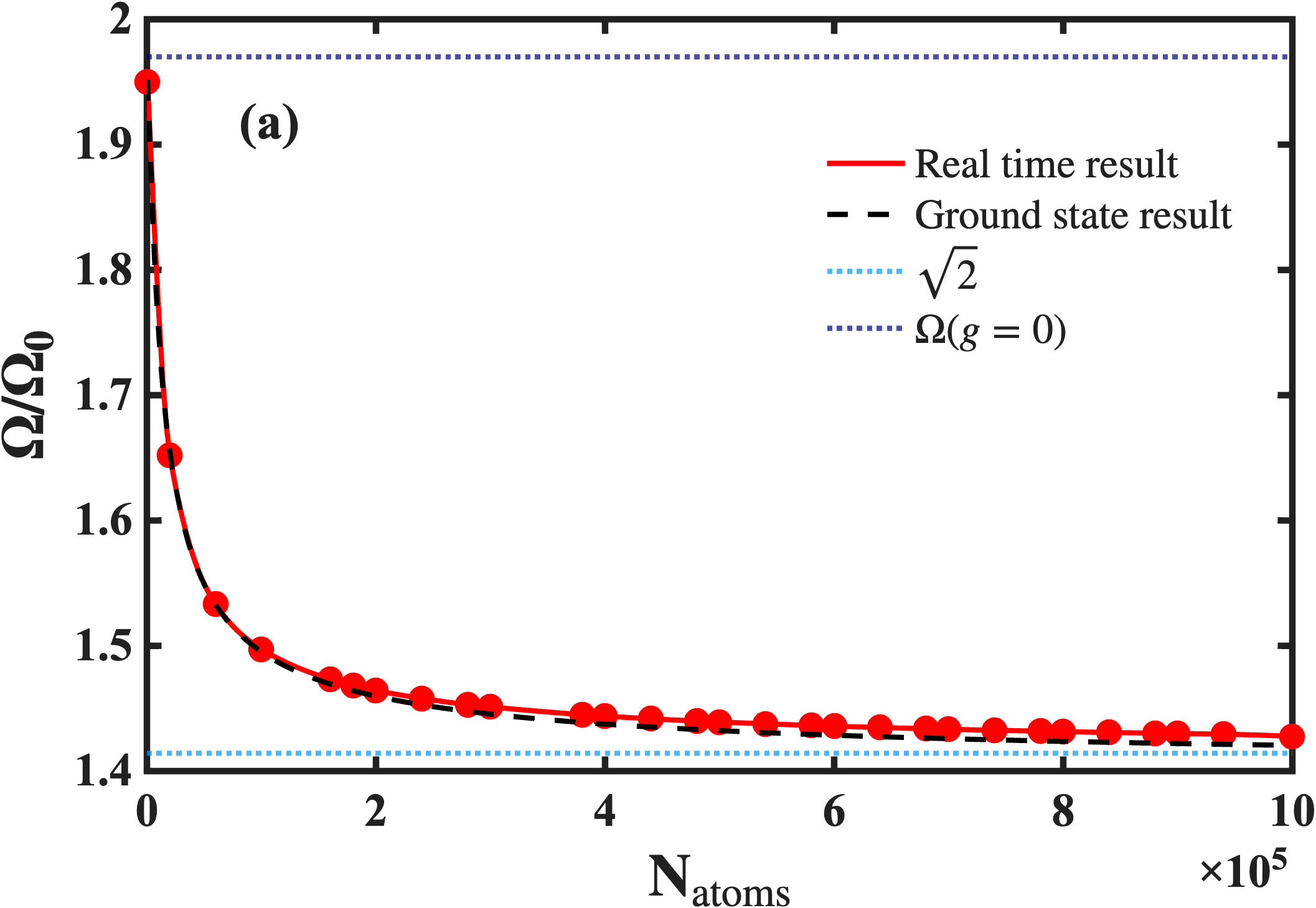}
\vspace{0.45em}
\includegraphics[width=\linewidth]{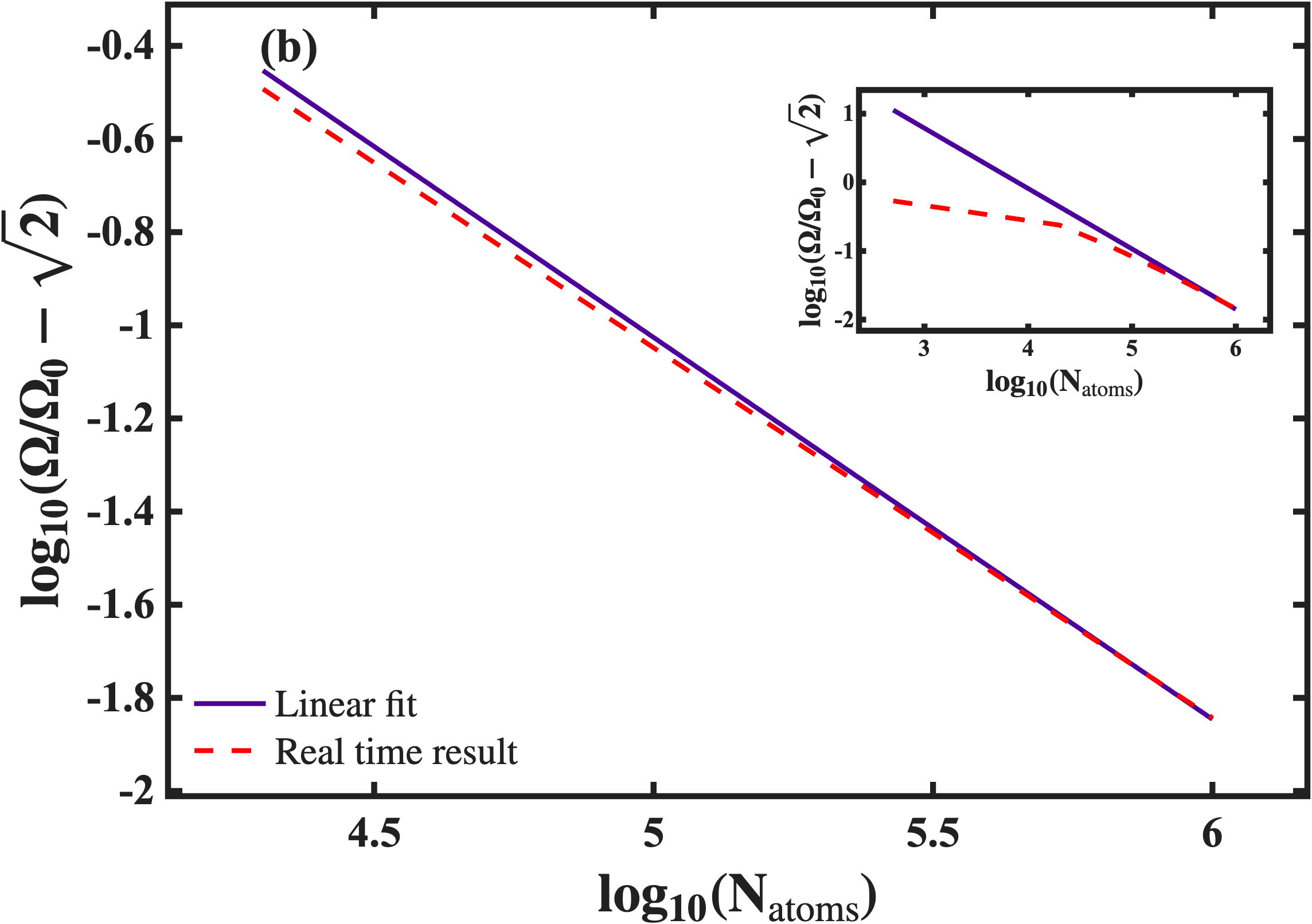}

\caption{Scissors mode oscillation frequency as a function of atom number for the LHY system. The upper panel shows the frequency extracted from numerical calculations (solid black curve; `Real time result') and the prediction of Eq.~(\ref{Frequency}) (red dashed curve; `Ground state result'). 
Both curves approach the infinite atom limit of $\sqrt{2}\Omega_0$ and the non-interacting limit, $\Omega(g=0)=\omega_x+\omega_y$ (dotted lines). The lower panel shows the data from the upper panel in the limit of large $N_{\textrm{atoms}}$ together with the linear (on the log-log plot) fit. The inset shows the deviation from the linear fit for smaller values of atoms.  
All results are obtained with $\omega_z=2\pi\times 110$~Hz, $\omega_x=\sqrt{1.3}\,\omega_z$, $\omega_y=\sqrt{0.7}\, \omega_z$, $a_{12}=100$ $a_0$, and the mass was taken to be the mass of rubidium atoms.  }
\label{fig:1}
\end{figure}

It is worth noting that $\langle xy\rangle$ provides an alternative way for the derivation of the frequency of the breathing mode~\cite{GuryOdelin1999}. Indeed, using the continuity equation, one can derive the equation for the evolution of $\langle xy\rangle$ in the linear approximation
 \begin{equation}
	 	\frac{\mathrm{d}^2\langle xy\rangle}{\mathrm{d}t^2} = -\frac{1}{N}\int \mathrm{d}\bm{r}~n_0(\bm{r})(\dot{v}_xy + \dot{v}_yx),
	 \end{equation}
which together with Eq.~(\ref{Simplified-Hydrodynamic}) confirms that $\langle xy\rangle$ is indeed an oscillatory function. Furthermore, one can derive a useful expression for the frequency 
 \begin{equation}
	 	\Omega^2 = (\omega_x^2-\omega_y^2)\frac{\int d\bm{r}~n_0(\bm{r})(y^2 + x^2)}{\int d\bm{r}~n_0(\bm{r})\big(y^2 - x^2\big)},
	 	\label{Frequency}
	 \end{equation}
     which agrees with the results above. In particular, it can be derived directly from Eq.~(\ref{eq:density_necessary}). The parameter $\langle y^2+x^2\rangle/\langle y^2-x^2\rangle$ is the deformation of the atomic cloud in the $xy$-plane. It defines the moment of inertia in terms of that of a rigid body~\cite{Pitaevskii2016}. This observation allows one to connect the frequency of the scissors mode to the moment of inertia of the system~\cite{GuryOdelin1999}. 

\begin{figure}
\centering
\includegraphics[width=0.45\textwidth]{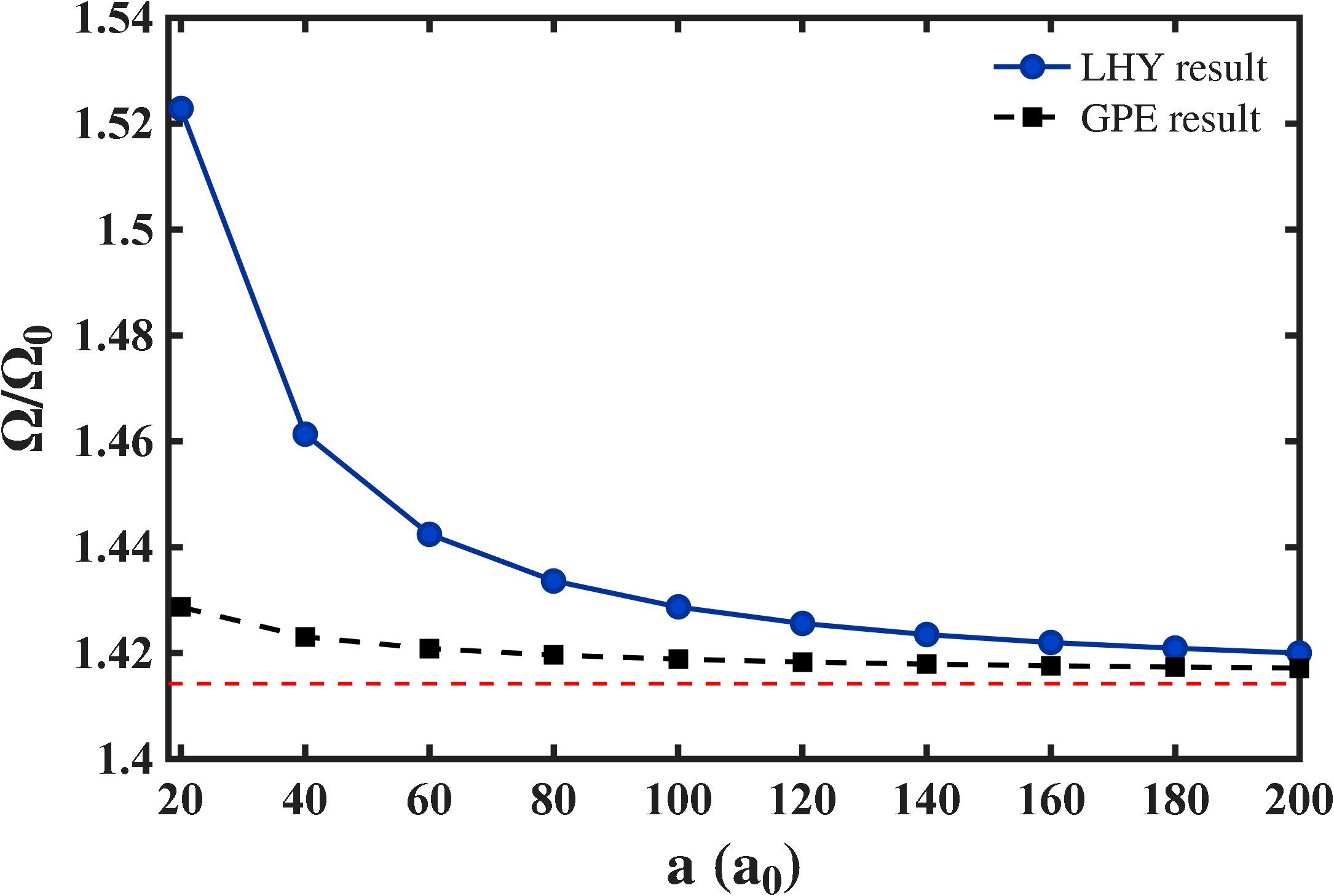}
\caption{ Scissors mode frequency as a function of the scattering length for a GPE system of 200000 atoms (black curve with square points) and an LHY system of one million atoms (green curve with circle points). A horizontal line at the value of $\sqrt{2}$ guides the eye to the large atom limit. 
The results are obtained with $\omega_z=2\pi\times 110$~Hz, $\omega_x=\sqrt{1.3}\,\omega_z$, $\omega_y=\sqrt{0.7}\, \omega_z$.}
\label{fig:4}
\end{figure}

\subsection{Numerical results for the LHY liquid.}
The derivations presented above rely on a linear approximation. To explore the scissors mode beyond this regime, we now turn to a numerical study of the system. The aim of this subsection is to identify and understand deviations from the scissors‑mode behavior predicted by our analytical results. We focus specifically on the LHY liquid ($\gamma=3$), which remains far less studied than the standard Gross–Pitaevskii equation. However, for the convenience of the reader, we present the frequency of the breathing mode as well as the dynamics of $\langle xy\rangle$ for the standard GPE ($\gamma=2$) in the Appendix. 

To carry out the numerical calculations, we solve the generalized GPE using the split‑step Crank–Nicolson method. Our numerical code is based on a modification of the open‑source software described in Ref.~\cite{kumar2015}, which we have adapted in our previous work~\cite{Shukla2024}. {\color{black} Some additional calculations were performed using standard split-step Fourier method.} 
Our numerical simulations are performed for ${}^{87}$Rb atoms assuming $\omega_z=2\pi\times 110$~Hz, $\omega_x=\sqrt{1.3}\,\omega_z$, $\omega_y=\sqrt{0.7}\, \omega_z$, and $a_{12}=100$ $a_0$, where $a_0$ is the Bohr radius. {\color{black}To present our results in a dimensionless form, we use the frequency $\Omega_0=
\omega_z$ to set the units. The corresponding unit of time, $T_{\mathrm{trap}} = \frac{2\pi}{\Omega_0
}$, is $T_{\mathrm{trap}} \approx 9.09 \textrm{ms}$.}  These parameters represent some generic cold-atom experiment without having a particular experimental setup in mind. 

\begin{figure*}
\centering
\includegraphics[width=\linewidth]{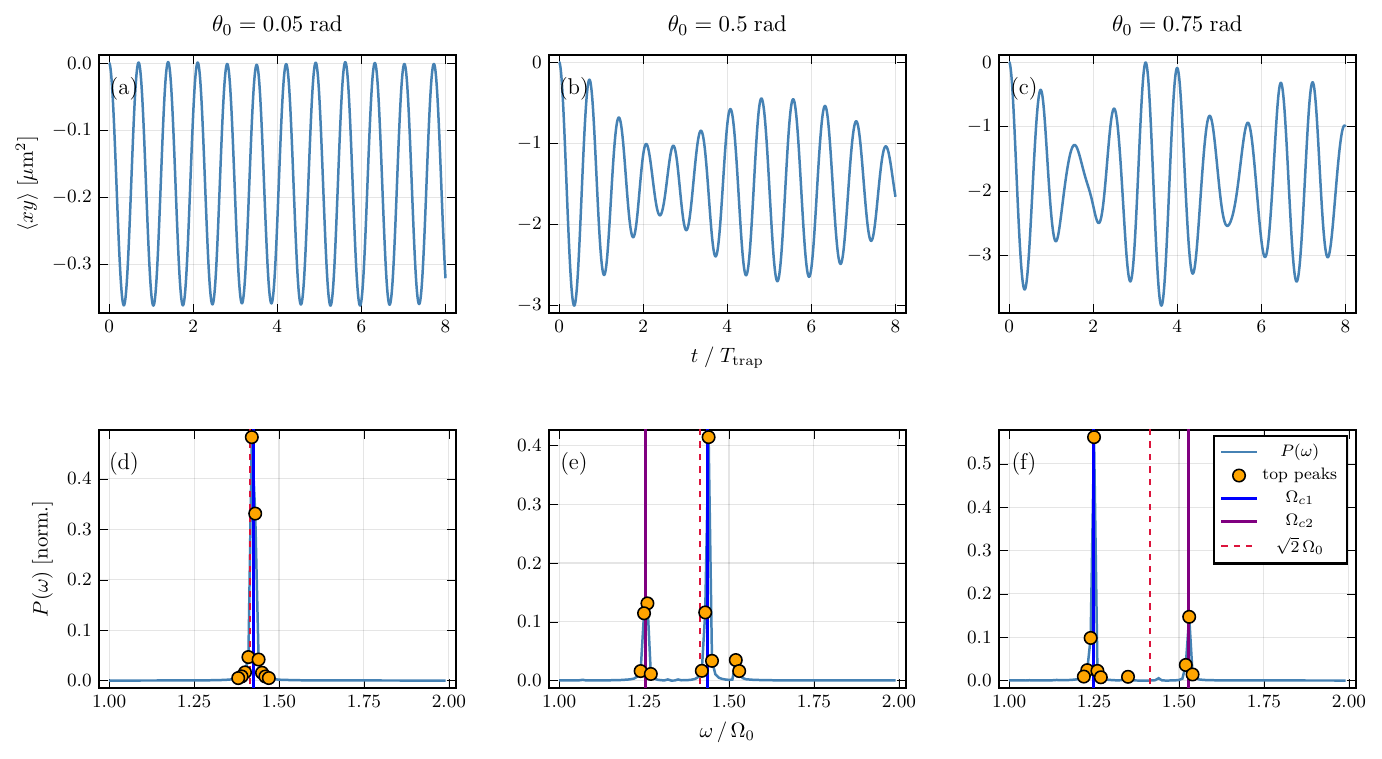}
{\color{black}
\caption{(a)-(c) Time dynamics of $\langle xy \rangle$ as a function of time for three different initial angles ($\theta_0$).  The numerical calculations were done for the LHY system with one million atoms, all other parameters are as in Fig.~\ref{fig:1}.  (d)-(f) Frequencies of the oscillations shown in panels (a)–(c). The filled circles indicate the ten largest frequency components obtained from the discrete Fourier transform of the numerical data. The solid curves are added to guide the eye.}
\label{fig:3}}
\end{figure*}

\textit{Dependence of the frequency on the interaction strength.} We begin by calculating the oscillation frequency of $\langle xy \rangle$ for $\theta_0\to0$ as a function of atom number, focusing specifically on the transition from the non-interacting regime to the strongly interacting (Thomas-Fermi) regime. Surprisingly, we find that for all interaction strengths, the oscillations of $\langle xy \rangle$ can be exceptionally well described by a single harmonic function. The quality of this single-harmonic fit is confirmed by a reduced $\chi^2$ per data point of less than 0.1\%. The extracted frequency is plotted in Fig.~\ref{fig:1}a) together with the result of Eq.~(\ref{Frequency}). Note the good agreement between the two results. In Fig.~\ref{fig:1}b), we present a subset of the data that illustrates the approach to the Thomas-Fermi limit. We observe 
$\Omega\sim 1/N^\alpha_{\textrm{atoms}}$ with $\alpha\simeq 0.6$. Unfortunately, we were unable to derive this result using standard boundary-layer methods~\cite{Fetter1998}, due to the significant anisotropy inherent in our system. In general, we observe that the number of atoms required to reach the Thomas-Fermi limit is greater for the LHY liquid than for the standard GPE (see Appendix). This difference is expected, as the LHY liquid exhibits weaker interactions at the same densities for the trapping potential considered, where the parameter $na^3$ remains small.

The dependence on the atom number translates to a dependence on the scattering length. We illustrate this in Fig.~\ref{fig:4}, which shows the frequency as a function of the scattering length for a fixed number of atoms. This figure mirrors the structure of Fig.~\ref{fig:1} and explicitly demonstrates that the standard GPE result remains closer to the Thomas-Fermi limit, even when the number of atoms in the LHY liquid is five times greater than that of the GPE system.

\textit{Persistence of scissors mode for large values of $\theta_0$}. Here, we investigate the oscillations of $\langle xy \rangle$ for various values of $\theta_0$ in the Thomas-Fermi limit, as shown in Fig.~\ref{fig:3}. For small perturbations of the trap, the time dynamics of $\langle xy \rangle$ are dominated by a single frequency. However, for larger values of $\theta_0$, at least two distinct modes must be included to accurately capture the dynamics.
Although the resulting behavior resembles beating, it is not associated with rotational flow, as observed in thermal clouds~\cite{Marag2000}. Instead, this phenomenon reflects the excitation of  additional modes of the condensate by the applied protocol.

{\color{black}To investigate these modes, we compute the time evolution over approximately 100 oscillations, yielding sufficient data to perform a fast Fourier transform. We then identify the ten largest frequency components, which group into two distinct clusters (see Fig.~\ref{fig:3}).
We find that, for angles $\theta_0 \lesssim 0.8$, the majority of the spectral weight ($\gtrsim 80\%$) is concentrated in these two clusters.

For $\theta_0 < 0.5$, the scissors-mode frequency is found to be nearly independent of $\theta_0$. The population of this mode exceeds $50\%$, which facilitates its experimental determination even under substantial trap deformations. We identify the second mode as a quadrupole mode. Indeed, the quadrupole mode is expected to dominate the dynamics at large angles, particularly at $\theta_0 = \pi/2$ (cf. Ref.~\cite{GuryOdelin1999}). Furthermore, its frequency lies close to that of known quadrupole modes. For instance, in the isotropic case $\omega_x = \omega_y = \omega_z$, the quadrupole mode has frequency $\Omega = \sqrt{2}\omega_z$~\cite{Heiselberg2004}.}

\section{Conclusion and Outlook}
\label{sec:conclusion}

This work presents a systematic analysis of scissors modes in systems governed by the generalized Gross-Pitaevskii equation, revealing that the frequency of the scissors mode is universal in the Thomas-Fermi regime and independent of the specific form of the nonlinearity. Furthermore, the study explores scissors modes in LHY liquids numerically, uncovering pronounced finite-size effects that should be important in current cold-atom setups.

From a theoretical standpoint, an intriguing avenue for future research lies in exploring scissors-like dynamics induced not by the geometry of the external trapping potential, but rather by anisotropic impurities embedded within the system. Such impurities could manifest as localized molecular dopants, for which a GPE-like framework has been developed~\cite{Suchorowski2025a}. Investigating this scenario holds the potential  to 
bridge cold-atom physics with the well-established field of helium droplet dynamics~\cite{Toennies2001}, by identifying universal collective excitations in quantum fluids.

Experimentally, this work suggests interesting avenues both in the standard GPE and LHY regime. In the GPE regime, previous experiments have only explored the regime of small oscillations~\cite{Marag2000, marago_moment_2002, Cozzini2003, Modugno2003}, at moderate scattering lengths. Thus, they served as a simultaneous probe of BEC and superfluidity. Experiments with adjustable scattering length could extend these investigations to probe superfluidiuty at weak interactions. Moreover, such experiments should be able to probe higher excitation frequencies at larger rotation angles. Finally, a new generation of heteronuclear LHY systems~\cite{Errico2019,Guo2021} with small losses will enable probing of the LHY regime. These systems are typically restricted to fixed scattering lengths and atom numbers to fulfill the LHY condition, but they should be able to explore the beating dynamics demonstrated in Fig.~\ref{fig:3}.



\section*{Acknowledgements}
The authors acknowledge that this material is based upon work supported by the National Science Foundation/EPSCoR RII Track-1: Emergent Quantum Materials and Technologies (EQUATE), Award OIA-2044049. The research has been also supported in part by the Novo Nordisk Foundation (grant reference number NNF25OC0102659).
J.A. acknowledges support from the Danish National Research Foundation through the Center of Excellence "CCQ" (DNRF152) and from the Novo Nordisk Foundation NERD grant (Grant No. NNF22OC0075986).

\appendix
\clearpage

\section{Results for the standard GPE}

This appendix collects the results of the GPE system. Scissors-mode behavior has been well-studied previously in GPE type condensates, at least in the Thomas-Fermi regime~\cite{GuryOdelin1999}, so we show our results here for the reader to compare quickly with our LHY results.  The behavior is mostly similar except that the GPE system reaches the large-N limit faster because the corresponding interactions are stronger than those in the LHY system. For instance, we observe that the scissors mode frequency reaches the universal value with an accuracy of 1\% already at $N_{\mathrm{atoms}}\simeq 10^5$.

\begin{figure}
\centering
\includegraphics[width=\linewidth]{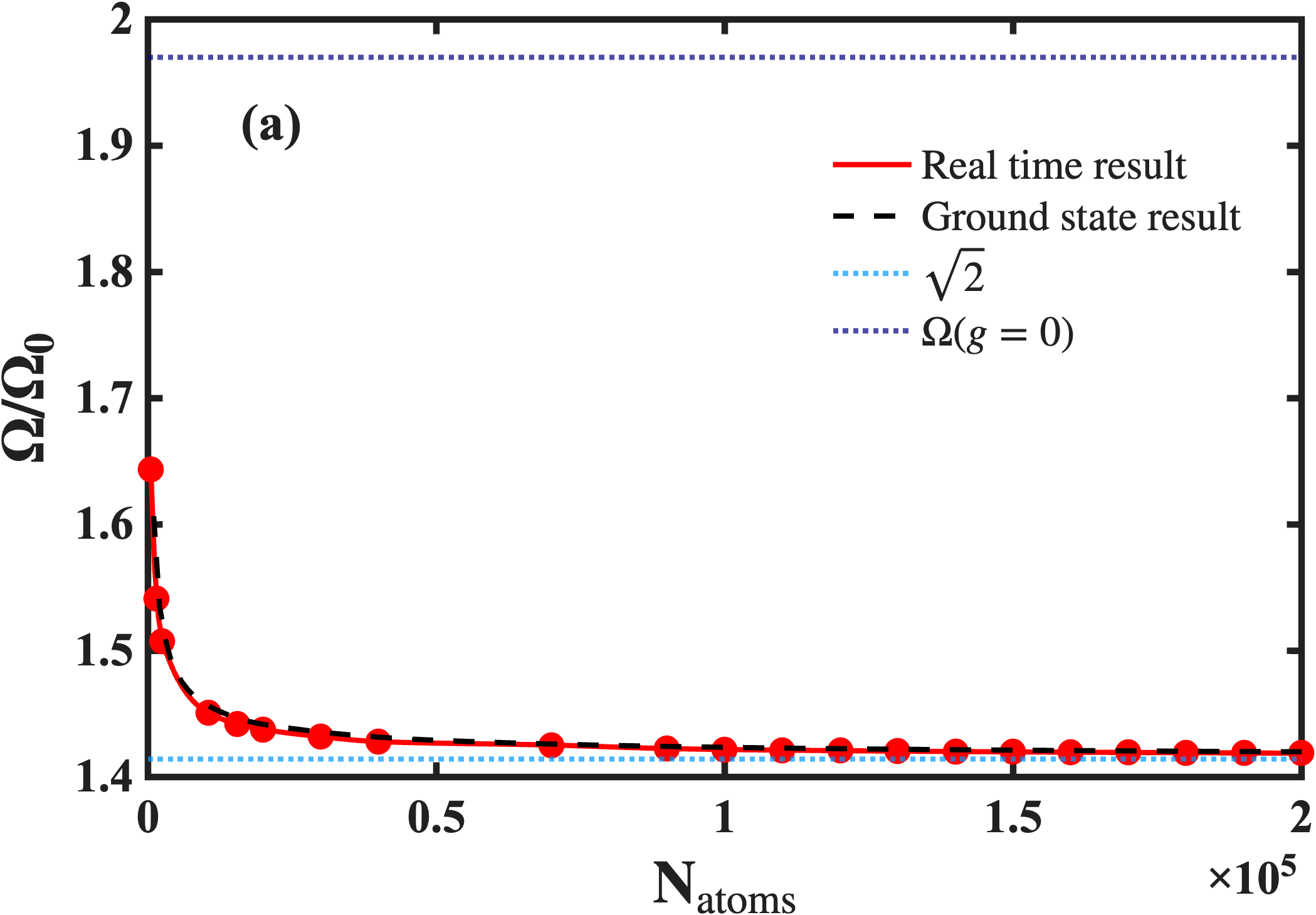}
\vspace{0.45em}
\includegraphics[width=\linewidth]{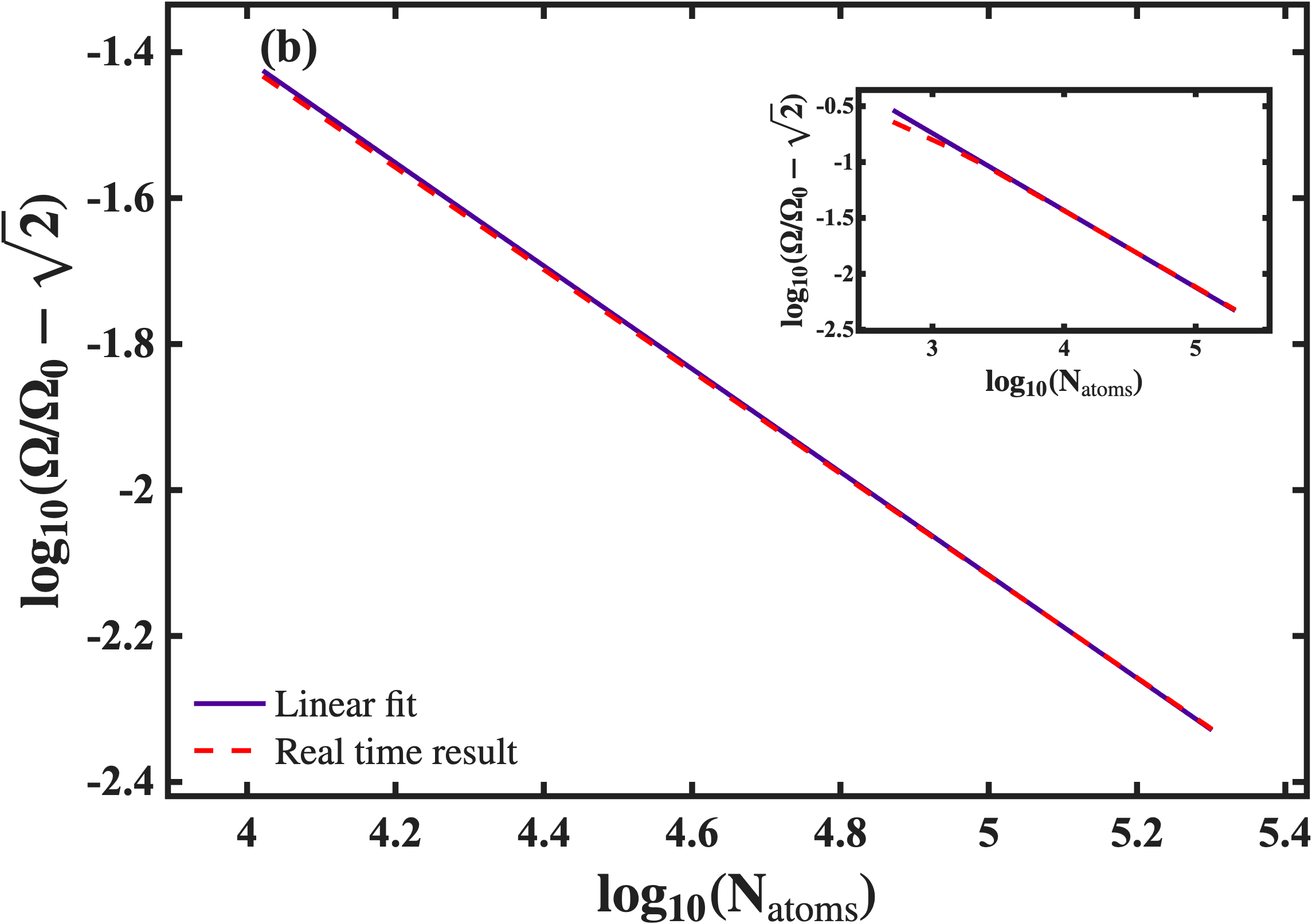}

\caption{Panel (a): Plot of the scissors mode frequency as a function of the number of atoms.  The small and large atom limits are shown as horizontal lines. Panel (b): A log-log plot where the abscissa is the logarithm of the number of atoms and the ordinate is the logarithm of the difference of the scissors mode frequency and $\sqrt{2}$, which is the large atom limit. These calculations were done for the same trap parameters as Fig \ref{fig:1}, and the scattering length was 100 $a_0$.}
\label{fig:2}
\end{figure}

\begin{figure*}
\centering
\includegraphics[width=\linewidth]{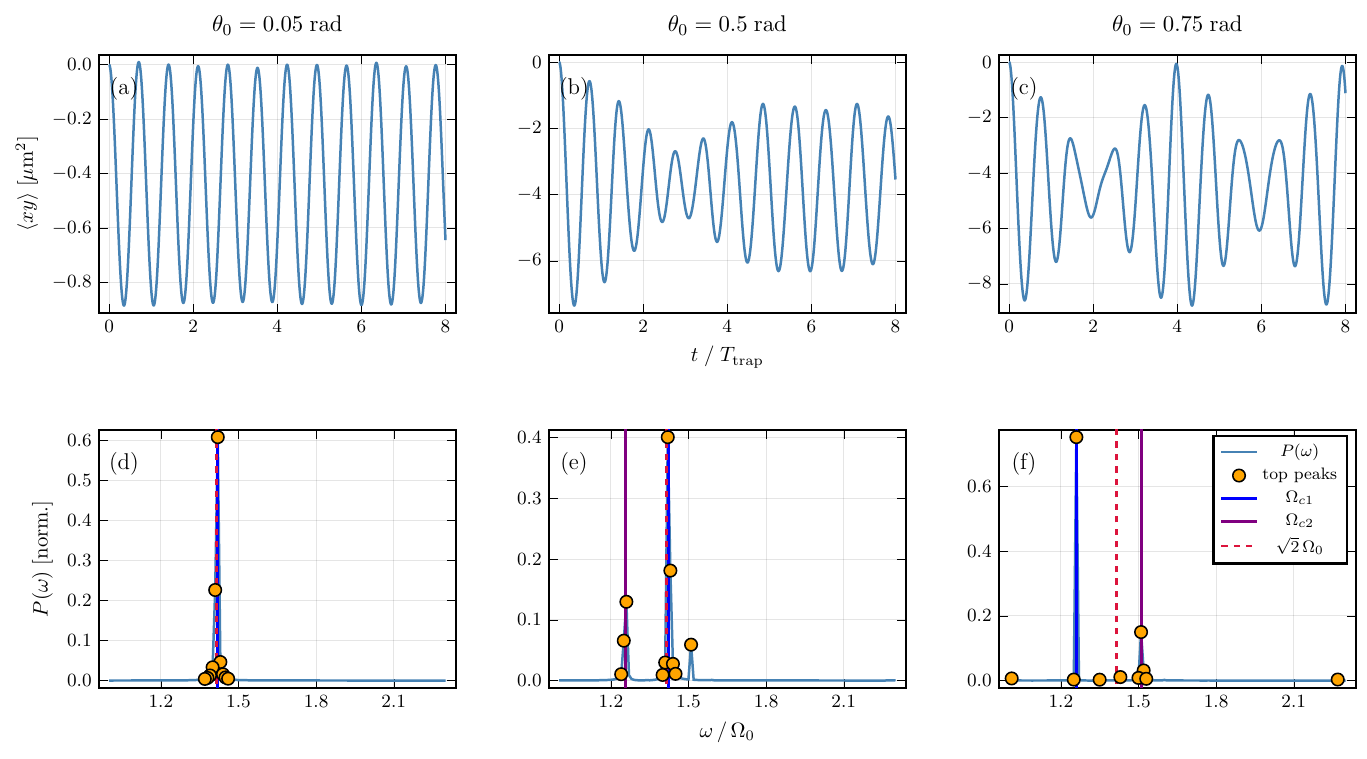}
\caption{ {\color{black}
(a)-(c) Time dynamics of $\langle xy \rangle$ as a function of time for three different initial angles ($\theta_0$).  The numerical calculations were done for the Gross-Pitaevskii equation with one million atoms, all other parameters are as in Fig.~\ref{fig:2}.  (d)-(f) Frequencies of the oscillations shown in panels (a)–(c). The filled circles indicate the ten largest frequency components obtained from the discrete Fourier transform of the numerical data. The solid curves are added to guide the eye.}
\label{fig:6}}
\end{figure*}

\clearpage
\bibliographystyle{apsrev4-1}
\bibliography{ref}

\begin{thebibliography}{36}%
\makeatletter
\providecommand \@ifxundefined [1]{%
 \@ifx{#1\undefined}
}%
\providecommand \@ifnum [1]{%
 \ifnum #1\expandafter \@firstoftwo
 \else \expandafter \@secondoftwo
 \fi
}%
\providecommand \@ifx [1]{%
 \ifx #1\expandafter \@firstoftwo
 \else \expandafter \@secondoftwo
 \fi
}%
\providecommand \natexlab [1]{#1}%
\providecommand \enquote  [1]{``#1''}%
\providecommand \bibnamefont  [1]{#1}%
\providecommand \bibfnamefont [1]{#1}%
\providecommand \citenamefont [1]{#1}%
\providecommand \href@noop [0]{\@secondoftwo}%
\providecommand \href [0]{\begingroup \@sanitize@url \@href}%
\providecommand \@href[1]{\@@startlink{#1}\@@href}%
\providecommand \@@href[1]{\endgroup#1\@@endlink}%
\providecommand \@sanitize@url [0]{\catcode `\\12\catcode `\$12\catcode
  `\&12\catcode `\#12\catcode `\^12\catcode `\_12\catcode `\%12\relax}%
\providecommand \@@startlink[1]{}%
\providecommand \@@endlink[0]{}%
\providecommand \url  [0]{\begingroup\@sanitize@url \@url }%
\providecommand \@url [1]{\endgroup\@href {#1}{\urlprefix }}%
\providecommand \urlprefix  [0]{URL }%
\providecommand \Eprint [0]{\href }%
\providecommand \doibase [0]{http://dx.doi.org/}%
\providecommand \selectlanguage [0]{\@gobble}%
\providecommand \bibinfo  [0]{\@secondoftwo}%
\providecommand \bibfield  [0]{\@secondoftwo}%
\providecommand \translation [1]{[#1]}%
\providecommand \BibitemOpen [0]{}%
\providecommand \bibitemStop [0]{}%
\providecommand \bibitemNoStop [0]{.\EOS\space}%
\providecommand \EOS [0]{\spacefactor3000\relax}%
\providecommand \BibitemShut  [1]{\csname bibitem#1\endcsname}%
\let\auto@bib@innerbib\@empty
\bibitem [{\citenamefont {Kartashov}\ \emph {et~al.}(2019)\citenamefont
  {Kartashov}, \citenamefont {Astrakharchik}, \citenamefont {Malomed},\ and\
  \citenamefont {Torner}}]{Kartashov2019}%
  \BibitemOpen
  \bibfield  {author} {\bibinfo {author} {\bibfnamefont {Y.~V.}\ \bibnamefont
  {Kartashov}}, \bibinfo {author} {\bibfnamefont {G.~E.}\ \bibnamefont
  {Astrakharchik}}, \bibinfo {author} {\bibfnamefont {B.~A.}\ \bibnamefont
  {Malomed}}, \ and\ \bibinfo {author} {\bibfnamefont {L.}~\bibnamefont
  {Torner}},\ }\href {\doibase 10.1038/s42254-019-0025-7} {\bibfield  {journal}
  {\bibinfo  {journal} {Nature Reviews Physics}\ }\textbf {\bibinfo {volume}
  {1}},\ \bibinfo {pages} {185–197} (\bibinfo {year} {2019})}\BibitemShut
  {NoStop}%
\bibitem [{\citenamefont {Dalfovo}\ \emph {et~al.}(1999)\citenamefont
  {Dalfovo}, \citenamefont {Giorgini}, \citenamefont {Pitaevskii},\ and\
  \citenamefont {Stringari}}]{Dalfovo1999}%
  \BibitemOpen
  \bibfield  {author} {\bibinfo {author} {\bibfnamefont {F.}~\bibnamefont
  {Dalfovo}}, \bibinfo {author} {\bibfnamefont {S.}~\bibnamefont {Giorgini}},
  \bibinfo {author} {\bibfnamefont {L.~P.}\ \bibnamefont {Pitaevskii}}, \ and\
  \bibinfo {author} {\bibfnamefont {S.}~\bibnamefont {Stringari}},\ }\href
  {\doibase 10.1103/RevModPhys.71.463} {\bibfield  {journal} {\bibinfo
  {journal} {Rev. Mod. Phys.}\ }\textbf {\bibinfo {volume} {71}},\ \bibinfo
  {pages} {463} (\bibinfo {year} {1999})}\BibitemShut {NoStop}%
\bibitem [{\citenamefont {Gross}(1961)}]{Gross1961}%
  \BibitemOpen
  \bibfield  {author} {\bibinfo {author} {\bibfnamefont {E.~P.}\ \bibnamefont
  {Gross}},\ }\href {\doibase 10.1007/bf02731494} {\bibfield  {journal}
  {\bibinfo  {journal} {Il Nuovo Cimento}\ }\textbf {\bibinfo {volume} {20}},\
  \bibinfo {pages} {454–477} (\bibinfo {year} {1961})}\BibitemShut {NoStop}%
\bibitem [{\citenamefont {Pitaevskii}(1961)}]{Pitaevskii1961}%
  \BibitemOpen
  \bibfield  {author} {\bibinfo {author} {\bibfnamefont {L.}~\bibnamefont
  {Pitaevskii}},\ }\href@noop {} {\bibfield  {journal} {\bibinfo  {journal}
  {Soviet Phys}\ }\textbf {\bibinfo {volume} {13}},\ \bibinfo {pages} {451}
  (\bibinfo {year} {1961})}\BibitemShut {NoStop}%
\bibitem [{\citenamefont {Kolomeisky}\ \emph {et~al.}(2000)\citenamefont
  {Kolomeisky}, \citenamefont {Newman}, \citenamefont {Straley},\ and\
  \citenamefont {Qi}}]{Kolomeisky2000}%
  \BibitemOpen
  \bibfield  {author} {\bibinfo {author} {\bibfnamefont {E.~B.}\ \bibnamefont
  {Kolomeisky}}, \bibinfo {author} {\bibfnamefont {T.~J.}\ \bibnamefont
  {Newman}}, \bibinfo {author} {\bibfnamefont {J.~P.}\ \bibnamefont {Straley}},
  \ and\ \bibinfo {author} {\bibfnamefont {X.}~\bibnamefont {Qi}},\ }\href
  {\doibase 10.1103/PhysRevLett.85.1146} {\bibfield  {journal} {\bibinfo
  {journal} {Phys. Rev. Lett.}\ }\textbf {\bibinfo {volume} {85}},\ \bibinfo
  {pages} {1146} (\bibinfo {year} {2000})}\BibitemShut {NoStop}%
\bibitem [{\citenamefont {K\"ohler}(2002)}]{Thorsten2002}%
  \BibitemOpen
  \bibfield  {author} {\bibinfo {author} {\bibfnamefont {T.}~\bibnamefont
  {K\"ohler}},\ }\href {\doibase 10.1103/PhysRevLett.89.210404} {\bibfield
  {journal} {\bibinfo  {journal} {Phys. Rev. Lett.}\ }\textbf {\bibinfo
  {volume} {89}},\ \bibinfo {pages} {210404} (\bibinfo {year}
  {2002})}\BibitemShut {NoStop}%
\bibitem [{\citenamefont {Pieri}\ and\ \citenamefont
  {Strinati}(2003)}]{Pieri2003}%
  \BibitemOpen
  \bibfield  {author} {\bibinfo {author} {\bibfnamefont {P.}~\bibnamefont
  {Pieri}}\ and\ \bibinfo {author} {\bibfnamefont {G.~C.}\ \bibnamefont
  {Strinati}},\ }\href {\doibase 10.1103/PhysRevLett.91.030401} {\bibfield
  {journal} {\bibinfo  {journal} {Phys. Rev. Lett.}\ }\textbf {\bibinfo
  {volume} {91}},\ \bibinfo {pages} {030401} (\bibinfo {year}
  {2003})}\BibitemShut {NoStop}%
\bibitem [{\citenamefont {Heiselberg}(2004)}]{Heiselberg2004}%
  \BibitemOpen
  \bibfield  {author} {\bibinfo {author} {\bibfnamefont {H.}~\bibnamefont
  {Heiselberg}},\ }\href {\doibase 10.1103/physrevlett.93.040402} {\bibfield
  {journal} {\bibinfo  {journal} {Physical Review Letters}\ }\textbf {\bibinfo
  {volume} {93}} (\bibinfo {year} {2004}),\
  10.1103/physrevlett.93.040402}\BibitemShut {NoStop}%
\bibitem [{\citenamefont {Choi}\ \emph {et~al.}(2015)\citenamefont {Choi},
  \citenamefont {Dunjko}, \citenamefont {Zhang},\ and\ \citenamefont
  {Olshanii}}]{Choi2015}%
  \BibitemOpen
  \bibfield  {author} {\bibinfo {author} {\bibfnamefont {S.}~\bibnamefont
  {Choi}}, \bibinfo {author} {\bibfnamefont {V.}~\bibnamefont {Dunjko}},
  \bibinfo {author} {\bibfnamefont {Z.~D.}\ \bibnamefont {Zhang}}, \ and\
  \bibinfo {author} {\bibfnamefont {M.}~\bibnamefont {Olshanii}},\ }\href
  {\doibase 10.1103/PhysRevLett.115.115302} {\bibfield  {journal} {\bibinfo
  {journal} {Phys. Rev. Lett.}\ }\textbf {\bibinfo {volume} {115}},\ \bibinfo
  {pages} {115302} (\bibinfo {year} {2015})}\BibitemShut {NoStop}%
\bibitem [{\citenamefont {Petrov}(2015)}]{Petrov2015}%
  \BibitemOpen
  \bibfield  {author} {\bibinfo {author} {\bibfnamefont {D.}~\bibnamefont
  {Petrov}},\ }\href {\doibase 10.1103/physrevlett.115.155302} {\bibfield
  {journal} {\bibinfo  {journal} {Physical Review Letters}\ }\textbf {\bibinfo
  {volume} {115}} (\bibinfo {year} {2015}),\
  10.1103/physrevlett.115.155302}\BibitemShut {NoStop}%
\bibitem [{\citenamefont {J\o{}rgensen}\ \emph {et~al.}(2018)\citenamefont
  {J\o{}rgensen}, \citenamefont {Bruun},\ and\ \citenamefont
  {Arlt}}]{Jorgensen2018}%
  \BibitemOpen
  \bibfield  {author} {\bibinfo {author} {\bibfnamefont {N.~B.}\ \bibnamefont
  {J\o{}rgensen}}, \bibinfo {author} {\bibfnamefont {G.~M.}\ \bibnamefont
  {Bruun}}, \ and\ \bibinfo {author} {\bibfnamefont {J.~J.}\ \bibnamefont
  {Arlt}},\ }\href {\doibase 10.1103/PhysRevLett.121.173403} {\bibfield
  {journal} {\bibinfo  {journal} {Phys. Rev. Lett.}\ }\textbf {\bibinfo
  {volume} {121}},\ \bibinfo {pages} {173403} (\bibinfo {year}
  {2018})}\BibitemShut {NoStop}%
\bibitem [{\citenamefont {Suchorowski}\ \emph {et~al.}(2026)\citenamefont
  {Suchorowski}, \citenamefont {Brauneis}, \citenamefont {Hammer},
  \citenamefont {Tomza},\ and\ \citenamefont {Volosniev}}]{suchorowski2026}%
  \BibitemOpen
  \bibfield  {author} {\bibinfo {author} {\bibfnamefont {M.}~\bibnamefont
  {Suchorowski}}, \bibinfo {author} {\bibfnamefont {F.}~\bibnamefont
  {Brauneis}}, \bibinfo {author} {\bibfnamefont {H.-W.}\ \bibnamefont
  {Hammer}}, \bibinfo {author} {\bibfnamefont {M.}~\bibnamefont {Tomza}}, \
  and\ \bibinfo {author} {\bibfnamefont {A.~G.}\ \bibnamefont {Volosniev}},\
  }\href {https://arxiv.org/abs/2511.10115} {\enquote {\bibinfo {title}
  {Generalized gross-pitaevskii equation for 2d bosons with attractive
  interactions},}\ } (\bibinfo {year} {2026}),\ \Eprint
  {http://arxiv.org/abs/2511.10115} {arXiv:2511.10115 [cond-mat.quant-gas]}
  \BibitemShut {NoStop}%
\bibitem [{\citenamefont {Guéry-Odelin}\ and\ \citenamefont
  {Stringari}(1999)}]{GuryOdelin1999}%
  \BibitemOpen
  \bibfield  {author} {\bibinfo {author} {\bibfnamefont {D.}~\bibnamefont
  {Guéry-Odelin}}\ and\ \bibinfo {author} {\bibfnamefont {S.}~\bibnamefont
  {Stringari}},\ }\href {\doibase 10.1103/physrevlett.83.4452} {\bibfield
  {journal} {\bibinfo  {journal} {Physical Review Letters}\ }\textbf {\bibinfo
  {volume} {83}},\ \bibinfo {pages} {4452–4455} (\bibinfo {year}
  {1999})}\BibitemShut {NoStop}%
\bibitem [{\citenamefont {Maragò}\ \emph {et~al.}(2000)\citenamefont
  {Maragò}, \citenamefont {Hopkins}, \citenamefont {Arlt}, \citenamefont
  {Hodby}, \citenamefont {Hechenblaikner},\ and\ \citenamefont
  {Foot}}]{Marag2000}%
  \BibitemOpen
  \bibfield  {author} {\bibinfo {author} {\bibfnamefont {O.~M.}\ \bibnamefont
  {Maragò}}, \bibinfo {author} {\bibfnamefont {S.~A.}\ \bibnamefont
  {Hopkins}}, \bibinfo {author} {\bibfnamefont {J.}~\bibnamefont {Arlt}},
  \bibinfo {author} {\bibfnamefont {E.}~\bibnamefont {Hodby}}, \bibinfo
  {author} {\bibfnamefont {G.}~\bibnamefont {Hechenblaikner}}, \ and\ \bibinfo
  {author} {\bibfnamefont {C.~J.}\ \bibnamefont {Foot}},\ }\href {\doibase
  10.1103/physrevlett.84.2056} {\bibfield  {journal} {\bibinfo  {journal}
  {Physical Review Letters}\ }\textbf {\bibinfo {volume} {84}},\ \bibinfo
  {pages} {2056–2059} (\bibinfo {year} {2000})}\BibitemShut {NoStop}%
\bibitem [{\citenamefont {Iudice}\ and\ \citenamefont
  {Palumbo}(1978)}]{Iudice1978}%
  \BibitemOpen
  \bibfield  {author} {\bibinfo {author} {\bibfnamefont {N.~L.}\ \bibnamefont
  {Iudice}}\ and\ \bibinfo {author} {\bibfnamefont {F.}~\bibnamefont
  {Palumbo}},\ }\href {\doibase 10.1103/physrevlett.41.1532} {\bibfield
  {journal} {\bibinfo  {journal} {Physical Review Letters}\ }\textbf {\bibinfo
  {volume} {41}},\ \bibinfo {pages} {1532–1534} (\bibinfo {year}
  {1978})}\BibitemShut {NoStop}%
\bibitem [{\citenamefont {Richter}(1995)}]{Richter1995}%
  \BibitemOpen
  \bibfield  {author} {\bibinfo {author} {\bibfnamefont {A.}~\bibnamefont
  {Richter}},\ }\href {\doibase 10.1016/0146-6410(95)00022-b} {\bibfield
  {journal} {\bibinfo  {journal} {Progress in Particle and Nuclear Physics}\
  }\textbf {\bibinfo {volume} {34}},\ \bibinfo {pages} {261–284} (\bibinfo
  {year} {1995})}\BibitemShut {NoStop}%
\bibitem [{\citenamefont {Maragò}\ \emph {et~al.}(2001)\citenamefont
  {Maragò}, \citenamefont {Hechenblaikner}, \citenamefont {Hodby},
  \citenamefont {Hopkins},\ and\ \citenamefont {Foot}}]{marago_moment_2002}%
  \BibitemOpen
  \bibfield  {author} {\bibinfo {author} {\bibfnamefont {O.~M.}\ \bibnamefont
  {Maragò}}, \bibinfo {author} {\bibfnamefont {G.}~\bibnamefont
  {Hechenblaikner}}, \bibinfo {author} {\bibfnamefont {E.}~\bibnamefont
  {Hodby}}, \bibinfo {author} {\bibfnamefont {S.~A.}\ \bibnamefont {Hopkins}},
  \ and\ \bibinfo {author} {\bibfnamefont {C.~J.}\ \bibnamefont {Foot}},\
  }\href {\doibase 10.1088/0953-8984/14/3/305} {\bibfield  {journal} {\bibinfo
  {journal} {Journal of Physics: Condensed Matter}\ }\textbf {\bibinfo {volume}
  {14}},\ \bibinfo {pages} {343} (\bibinfo {year} {2001})}\BibitemShut
  {NoStop}%
\bibitem [{\citenamefont {Cozzini}\ \emph {et~al.}(2003)\citenamefont
  {Cozzini}, \citenamefont {Stringari}, \citenamefont {Bretin}, \citenamefont
  {Rosenbusch},\ and\ \citenamefont {Dalibard}}]{Cozzini2003}%
  \BibitemOpen
  \bibfield  {author} {\bibinfo {author} {\bibfnamefont {M.}~\bibnamefont
  {Cozzini}}, \bibinfo {author} {\bibfnamefont {S.}~\bibnamefont {Stringari}},
  \bibinfo {author} {\bibfnamefont {V.}~\bibnamefont {Bretin}}, \bibinfo
  {author} {\bibfnamefont {P.}~\bibnamefont {Rosenbusch}}, \ and\ \bibinfo
  {author} {\bibfnamefont {J.}~\bibnamefont {Dalibard}},\ }\href {\doibase
  10.1103/PhysRevA.67.021602} {\bibfield  {journal} {\bibinfo  {journal}
  {Physical Review A}\ }\textbf {\bibinfo {volume} {67}} (\bibinfo {year}
  {2003}),\ 10.1103/PhysRevA.67.021602}\BibitemShut {NoStop}%
\bibitem [{\citenamefont {Modugno}\ \emph {et~al.}(2003)\citenamefont
  {Modugno}, \citenamefont {Modugno}, \citenamefont {Roati}, \citenamefont
  {Fort},\ and\ \citenamefont {Inguscio}}]{Modugno2003}%
  \BibitemOpen
  \bibfield  {author} {\bibinfo {author} {\bibfnamefont {M.}~\bibnamefont
  {Modugno}}, \bibinfo {author} {\bibfnamefont {G.}~\bibnamefont {Modugno}},
  \bibinfo {author} {\bibfnamefont {G.}~\bibnamefont {Roati}}, \bibinfo
  {author} {\bibfnamefont {C.}~\bibnamefont {Fort}}, \ and\ \bibinfo {author}
  {\bibfnamefont {M.}~\bibnamefont {Inguscio}},\ }\href {\doibase
  10.1103/PhysRevA.67.023608} {\bibfield  {journal} {\bibinfo  {journal}
  {Physical Review A}\ }\textbf {\bibinfo {volume} {67}} (\bibinfo {year}
  {2003}),\ 10.1103/PhysRevA.67.023608}\BibitemShut {NoStop}%
\bibitem [{\citenamefont {van Bijnen}\ \emph {et~al.}(2010)\citenamefont {van
  Bijnen}, \citenamefont {Parker}, \citenamefont {Kokkelmans}, \citenamefont
  {Martin},\ and\ \citenamefont {O’Dell}}]{vanBijnen2010}%
  \BibitemOpen
  \bibfield  {author} {\bibinfo {author} {\bibfnamefont {R.~M.~W.}\
  \bibnamefont {van Bijnen}}, \bibinfo {author} {\bibfnamefont {N.~G.}\
  \bibnamefont {Parker}}, \bibinfo {author} {\bibfnamefont {S.~J. J. M.~F.}\
  \bibnamefont {Kokkelmans}}, \bibinfo {author} {\bibfnamefont {A.~M.}\
  \bibnamefont {Martin}}, \ and\ \bibinfo {author} {\bibfnamefont {D.~H.~J.}\
  \bibnamefont {O’Dell}},\ }\href {\doibase 10.1103/physreva.82.033612}
  {\bibfield  {journal} {\bibinfo  {journal} {Physical Review A}\ }\textbf
  {\bibinfo {volume} {82}} (\bibinfo {year} {2010}),\
  10.1103/physreva.82.033612}\BibitemShut {NoStop}%
\bibitem [{\citenamefont {Ferrier-Barbut}\ \emph {et~al.}(2018)\citenamefont
  {Ferrier-Barbut}, \citenamefont {Wenzel}, \citenamefont {B\"{o}ttcher},
  \citenamefont {Langen}, \citenamefont {Isoard}, \citenamefont {Stringari},\
  and\ \citenamefont {Pfau}}]{FerrierBarbut2018}%
  \BibitemOpen
  \bibfield  {author} {\bibinfo {author} {\bibfnamefont {I.}~\bibnamefont
  {Ferrier-Barbut}}, \bibinfo {author} {\bibfnamefont {M.}~\bibnamefont
  {Wenzel}}, \bibinfo {author} {\bibfnamefont {F.}~\bibnamefont
  {B\"{o}ttcher}}, \bibinfo {author} {\bibfnamefont {T.}~\bibnamefont
  {Langen}}, \bibinfo {author} {\bibfnamefont {M.}~\bibnamefont {Isoard}},
  \bibinfo {author} {\bibfnamefont {S.}~\bibnamefont {Stringari}}, \ and\
  \bibinfo {author} {\bibfnamefont {T.}~\bibnamefont {Pfau}},\ }\href {\doibase
  10.1103/physrevlett.120.160402} {\bibfield  {journal} {\bibinfo  {journal}
  {Physical Review Letters}\ }\textbf {\bibinfo {volume} {120}} (\bibinfo
  {year} {2018}),\ 10.1103/physrevlett.120.160402}\BibitemShut {NoStop}%
\bibitem [{\citenamefont {Pitaevskii}\ and\ \citenamefont
  {Stringari}(2016)}]{Pitaevskii2016}%
  \BibitemOpen
  \bibfield  {author} {\bibinfo {author} {\bibfnamefont {L.~P.}\ \bibnamefont
  {Pitaevskii}}\ and\ \bibinfo {author} {\bibfnamefont {S.}~\bibnamefont
  {Stringari}},\ }\href {https://books.google.de/books?id=k\_ZGCwAAQBAJ} {\emph
  {\bibinfo {title} {Bose-Einstein Condensation and Superfluidity}}},\
  International series of monographs on physics\ (\bibinfo  {publisher} {Oxford
  University Press},\ \bibinfo {year} {2016})\BibitemShut {NoStop}%
\bibitem [{\citenamefont {Schmitt}\ \emph {et~al.}(2016)\citenamefont
  {Schmitt}, \citenamefont {Wenzel}, \citenamefont {Böttcher}, \citenamefont
  {Ferrier-Barbut},\ and\ \citenamefont {Pfau}}]{schmitt_self-bound_2016}%
  \BibitemOpen
  \bibfield  {author} {\bibinfo {author} {\bibfnamefont {M.}~\bibnamefont
  {Schmitt}}, \bibinfo {author} {\bibfnamefont {M.}~\bibnamefont {Wenzel}},
  \bibinfo {author} {\bibfnamefont {F.}~\bibnamefont {Böttcher}}, \bibinfo
  {author} {\bibfnamefont {I.}~\bibnamefont {Ferrier-Barbut}}, \ and\ \bibinfo
  {author} {\bibfnamefont {T.}~\bibnamefont {Pfau}},\ }\href {\doibase
  10.1038/nature20126} {\bibfield  {journal} {\bibinfo  {journal} {Nature}\
  }\textbf {\bibinfo {volume} {539}},\ \bibinfo {pages} {259} (\bibinfo {year}
  {2016})},\ \bibinfo {note} {publisher: Nature Publishing Group}\BibitemShut
  {NoStop}%
\bibitem [{\citenamefont {Cabrera}\ \emph {et~al.}(2018)\citenamefont
  {Cabrera}, \citenamefont {Tanzi}, \citenamefont {Sanz}, \citenamefont
  {Naylor}, \citenamefont {Thomas}, \citenamefont {Cheiney}, \citenamefont
  {Tarruell}, \citenamefont {Tarrue},\ and\ \citenamefont
  {Tarruell}}]{cabrera_quantum_2018}%
  \BibitemOpen
  \bibfield  {author} {\bibinfo {author} {\bibfnamefont {C.~R.}\ \bibnamefont
  {Cabrera}}, \bibinfo {author} {\bibfnamefont {L.}~\bibnamefont {Tanzi}},
  \bibinfo {author} {\bibfnamefont {J.}~\bibnamefont {Sanz}}, \bibinfo {author}
  {\bibfnamefont {B.}~\bibnamefont {Naylor}}, \bibinfo {author} {\bibfnamefont
  {P.}~\bibnamefont {Thomas}}, \bibinfo {author} {\bibfnamefont
  {P.}~\bibnamefont {Cheiney}}, \bibinfo {author} {\bibfnamefont
  {L.}~\bibnamefont {Tarruell}}, \bibinfo {author} {\bibfnamefont
  {L.}~\bibnamefont {Tarrue}}, \ and\ \bibinfo {author} {\bibfnamefont
  {L.}~\bibnamefont {Tarruell}},\ }\href {\doibase 10.1126/science.aao5686}
  {\bibfield  {journal} {\bibinfo  {journal} {Science}\ }\textbf {\bibinfo
  {volume} {359}},\ \bibinfo {pages} {301} (\bibinfo {year} {2018})},\ \bibinfo
  {note} {arXiv: 1708.07806v1 Publisher: American Association for the
  Advancement of Science}\BibitemShut {NoStop}%
\bibitem [{\citenamefont {Semeghini}\ \emph {et~al.}(2018)\citenamefont
  {Semeghini}, \citenamefont {Ferioli}, \citenamefont {Masi}, \citenamefont
  {Mazzinghi}, \citenamefont {Wolswijk}, \citenamefont {Minardi}, \citenamefont
  {Modugno}, \citenamefont {Modugno}, \citenamefont {Inguscio},\ and\
  \citenamefont {Fattori}}]{semeghini2018}%
  \BibitemOpen
  \bibfield  {author} {\bibinfo {author} {\bibfnamefont {G.}~\bibnamefont
  {Semeghini}}, \bibinfo {author} {\bibfnamefont {G.}~\bibnamefont {Ferioli}},
  \bibinfo {author} {\bibfnamefont {L.}~\bibnamefont {Masi}}, \bibinfo {author}
  {\bibfnamefont {C.}~\bibnamefont {Mazzinghi}}, \bibinfo {author}
  {\bibfnamefont {L.}~\bibnamefont {Wolswijk}}, \bibinfo {author}
  {\bibfnamefont {F.}~\bibnamefont {Minardi}}, \bibinfo {author} {\bibfnamefont
  {M.}~\bibnamefont {Modugno}}, \bibinfo {author} {\bibfnamefont
  {G.}~\bibnamefont {Modugno}}, \bibinfo {author} {\bibfnamefont
  {M.}~\bibnamefont {Inguscio}}, \ and\ \bibinfo {author} {\bibfnamefont
  {M.}~\bibnamefont {Fattori}},\ }\href {\doibase
  10.1103/PhysRevLett.120.235301} {\bibfield  {journal} {\bibinfo  {journal}
  {Physical Review Letters}\ }\textbf {\bibinfo {volume} {120}},\ \bibinfo
  {pages} {235301} (\bibinfo {year} {2018})},\ \bibinfo {note} {arXiv:
  1710.10890 Publisher: American Physical Society}\BibitemShut {NoStop}%
\bibitem [{\citenamefont {Skov}\ \emph {et~al.}(2021)\citenamefont {Skov},
  \citenamefont {Skou}, \citenamefont {Jørgensen},\ and\ \citenamefont
  {Arlt}}]{Skov2021}%
  \BibitemOpen
  \bibfield  {author} {\bibinfo {author} {\bibfnamefont {T.~G.}\ \bibnamefont
  {Skov}}, \bibinfo {author} {\bibfnamefont {M.~G.}\ \bibnamefont {Skou}},
  \bibinfo {author} {\bibfnamefont {N.~B.}\ \bibnamefont {Jørgensen}}, \ and\
  \bibinfo {author} {\bibfnamefont {J.~J.}\ \bibnamefont {Arlt}},\ }\href
  {\doibase 10.1103/physrevlett.126.230404} {\bibfield  {journal} {\bibinfo
  {journal} {Physical Review Letters}\ }\textbf {\bibinfo {volume} {126}}
  (\bibinfo {year} {2021}),\ 10.1103/physrevlett.126.230404}\BibitemShut
  {NoStop}%
\bibitem [{\citenamefont {Luo}\ \emph {et~al.}(2020)\citenamefont {Luo},
  \citenamefont {Pang}, \citenamefont {Liu}, \citenamefont {Li},\ and\
  \citenamefont {Malomed}}]{Luo2020}%
  \BibitemOpen
  \bibfield  {author} {\bibinfo {author} {\bibfnamefont {Z.-H.}\ \bibnamefont
  {Luo}}, \bibinfo {author} {\bibfnamefont {W.}~\bibnamefont {Pang}}, \bibinfo
  {author} {\bibfnamefont {B.}~\bibnamefont {Liu}}, \bibinfo {author}
  {\bibfnamefont {Y.-Y.}\ \bibnamefont {Li}}, \ and\ \bibinfo {author}
  {\bibfnamefont {B.~A.}\ \bibnamefont {Malomed}},\ }\href {\doibase
  10.1007/s11467-020-1020-2} {\bibfield  {journal} {\bibinfo  {journal}
  {Frontiers of Physics}\ }\textbf {\bibinfo {volume} {16}} (\bibinfo {year}
  {2020}),\ 10.1007/s11467-020-1020-2}\BibitemShut {NoStop}%
\bibitem [{\citenamefont {Pethick}\ and\ \citenamefont
  {Smith}(2008)}]{Pethick2008}%
  \BibitemOpen
  \bibfield  {author} {\bibinfo {author} {\bibfnamefont {C.~J.}\ \bibnamefont
  {Pethick}}\ and\ \bibinfo {author} {\bibfnamefont {H.}~\bibnamefont
  {Smith}},\ }\href {\doibase 10.1017/cbo9780511802850} {\emph {\bibinfo
  {title} {Bose–Einstein Condensation in Dilute Gases}}}\ (\bibinfo
  {publisher} {Cambridge University Press},\ \bibinfo {year}
  {2008})\BibitemShut {NoStop}%
\bibitem [{\citenamefont {Rossi}\ \emph {et~al.}(2016)\citenamefont {Rossi},
  \citenamefont {Dubessy}, \citenamefont {Merloti}, \citenamefont {Herve},
  \citenamefont {Badr}, \citenamefont {Perrin}, \citenamefont {Longchambon},\
  and\ \citenamefont {Perrin}}]{Rossi2016}%
  \BibitemOpen
  \bibfield  {author} {\bibinfo {author} {\bibfnamefont {C.~D.}\ \bibnamefont
  {Rossi}}, \bibinfo {author} {\bibfnamefont {R.}~\bibnamefont {Dubessy}},
  \bibinfo {author} {\bibfnamefont {K.}~\bibnamefont {Merloti}}, \bibinfo
  {author} {\bibfnamefont {M.~d. G.~d.}\ \bibnamefont {Herve}}, \bibinfo
  {author} {\bibfnamefont {T.}~\bibnamefont {Badr}}, \bibinfo {author}
  {\bibfnamefont {A.}~\bibnamefont {Perrin}}, \bibinfo {author} {\bibfnamefont
  {L.}~\bibnamefont {Longchambon}}, \ and\ \bibinfo {author} {\bibfnamefont
  {H.}~\bibnamefont {Perrin}},\ }\href {\doibase 10.1088/1367-2630/18/6/062001}
  {\bibfield  {journal} {\bibinfo  {journal} {New Journal of Physics}\ }\textbf
  {\bibinfo {volume} {18}},\ \bibinfo {pages} {062001} (\bibinfo {year}
  {2016})}\BibitemShut {NoStop}%
\bibitem [{\citenamefont {Kumar}\ \emph {et~al.}(2015)\citenamefont {Kumar},
  \citenamefont {Young-S}, \citenamefont {Vudragovi{\'c}}, \citenamefont
  {Bala{\v{z}}}, \citenamefont {Muruganandam},\ and\ \citenamefont
  {Adhikari}}]{kumar2015}%
  \BibitemOpen
  \bibfield  {author} {\bibinfo {author} {\bibfnamefont {R.~K.}\ \bibnamefont
  {Kumar}}, \bibinfo {author} {\bibfnamefont {L.~E.}\ \bibnamefont {Young-S}},
  \bibinfo {author} {\bibfnamefont {D.}~\bibnamefont {Vudragovi{\'c}}},
  \bibinfo {author} {\bibfnamefont {A.}~\bibnamefont {Bala{\v{z}}}}, \bibinfo
  {author} {\bibfnamefont {P.}~\bibnamefont {Muruganandam}}, \ and\ \bibinfo
  {author} {\bibfnamefont {S.~K.}\ \bibnamefont {Adhikari}},\ }\href@noop {}
  {\bibfield  {journal} {\bibinfo  {journal} {Computer Physics Communications}\
  }\textbf {\bibinfo {volume} {195}},\ \bibinfo {pages} {117} (\bibinfo {year}
  {2015})}\BibitemShut {NoStop}%
\bibitem [{\citenamefont {Shukla}\ \emph {et~al.}(2024)\citenamefont {Shukla},
  \citenamefont {Volosniev},\ and\ \citenamefont {Armstrong}}]{Shukla2024}%
  \BibitemOpen
  \bibfield  {author} {\bibinfo {author} {\bibfnamefont {N.}~\bibnamefont
  {Shukla}}, \bibinfo {author} {\bibfnamefont {A.~G.}\ \bibnamefont
  {Volosniev}}, \ and\ \bibinfo {author} {\bibfnamefont {J.~R.}\ \bibnamefont
  {Armstrong}},\ }\href {\doibase 10.1103/physreva.110.053317} {\bibfield
  {journal} {\bibinfo  {journal} {Physical Review A}\ }\textbf {\bibinfo
  {volume} {110}} (\bibinfo {year} {2024}),\
  10.1103/physreva.110.053317}\BibitemShut {NoStop}%
\bibitem [{\citenamefont {Fetter}\ and\ \citenamefont
  {Feder}(1998)}]{Fetter1998}%
  \BibitemOpen
  \bibfield  {author} {\bibinfo {author} {\bibfnamefont {A.~L.}\ \bibnamefont
  {Fetter}}\ and\ \bibinfo {author} {\bibfnamefont {D.~L.}\ \bibnamefont
  {Feder}},\ }\href {\doibase 10.1103/physreva.58.3185} {\bibfield  {journal}
  {\bibinfo  {journal} {Physical Review A}\ }\textbf {\bibinfo {volume} {58}},\
  \bibinfo {pages} {3185–3194} (\bibinfo {year} {1998})}\BibitemShut
  {NoStop}%
\bibitem [{\citenamefont {Suchorowski}\ \emph {et~al.}(2025)\citenamefont
  {Suchorowski}, \citenamefont {Badamshina}, \citenamefont {Lemeshko},
  \citenamefont {Tomza},\ and\ \citenamefont {Volosniev}}]{Suchorowski2025a}%
  \BibitemOpen
  \bibfield  {author} {\bibinfo {author} {\bibfnamefont {M.}~\bibnamefont
  {Suchorowski}}, \bibinfo {author} {\bibfnamefont {A.}~\bibnamefont
  {Badamshina}}, \bibinfo {author} {\bibfnamefont {M.}~\bibnamefont
  {Lemeshko}}, \bibinfo {author} {\bibfnamefont {M.}~\bibnamefont {Tomza}}, \
  and\ \bibinfo {author} {\bibfnamefont {A.~G.}\ \bibnamefont {Volosniev}},\
  }\href {\doibase 10.21468/scipostphys.18.2.059} {\bibfield  {journal}
  {\bibinfo  {journal} {SciPost Physics}\ }\textbf {\bibinfo {volume} {18}}
  (\bibinfo {year} {2025}),\ 10.21468/scipostphys.18.2.059}\BibitemShut
  {NoStop}%
\bibitem [{\citenamefont {Toennies}\ \emph {et~al.}(2001)\citenamefont
  {Toennies}, \citenamefont {Vilesov},\ and\ \citenamefont
  {Whaley}}]{Toennies2001}%
  \BibitemOpen
  \bibfield  {author} {\bibinfo {author} {\bibfnamefont {J.~P.}\ \bibnamefont
  {Toennies}}, \bibinfo {author} {\bibfnamefont {A.~F.}\ \bibnamefont
  {Vilesov}}, \ and\ \bibinfo {author} {\bibfnamefont {K.~B.}\ \bibnamefont
  {Whaley}},\ }\href {\doibase 10.1063/1.1359707} {\bibfield  {journal}
  {\bibinfo  {journal} {Physics Today}\ }\textbf {\bibinfo {volume} {54}},\
  \bibinfo {pages} {31–37} (\bibinfo {year} {2001})}\BibitemShut {NoStop}%
\bibitem [{\citenamefont {D'Errico}\ \emph {et~al.}(2019)\citenamefont
  {D'Errico}, \citenamefont {Burchianti}, \citenamefont {Prevedelli},
  \citenamefont {Salasnich}, \citenamefont {Ancilotto}, \citenamefont
  {Modugno}, \citenamefont {Minardi},\ and\ \citenamefont {Fort}}]{Errico2019}%
  \BibitemOpen
  \bibfield  {author} {\bibinfo {author} {\bibfnamefont {C.}~\bibnamefont
  {D'Errico}}, \bibinfo {author} {\bibfnamefont {A.}~\bibnamefont
  {Burchianti}}, \bibinfo {author} {\bibfnamefont {M.}~\bibnamefont
  {Prevedelli}}, \bibinfo {author} {\bibfnamefont {L.}~\bibnamefont
  {Salasnich}}, \bibinfo {author} {\bibfnamefont {F.}~\bibnamefont
  {Ancilotto}}, \bibinfo {author} {\bibfnamefont {M.}~\bibnamefont {Modugno}},
  \bibinfo {author} {\bibfnamefont {F.}~\bibnamefont {Minardi}}, \ and\
  \bibinfo {author} {\bibfnamefont {C.}~\bibnamefont {Fort}},\ }\href {\doibase
  10.1103/PhysRevResearch.1.033155} {\bibfield  {journal} {\bibinfo  {journal}
  {Phys. Rev. Res.}\ }\textbf {\bibinfo {volume} {1}},\ \bibinfo {pages}
  {033155} (\bibinfo {year} {2019})}\BibitemShut {NoStop}%
\bibitem [{\citenamefont {Guo}\ \emph {et~al.}(2021)\citenamefont {Guo},
  \citenamefont {Jia}, \citenamefont {Li}, \citenamefont {Ma}, \citenamefont
  {Hutson}, \citenamefont {Cui},\ and\ \citenamefont {Wang}}]{Guo2021}%
  \BibitemOpen
  \bibfield  {author} {\bibinfo {author} {\bibfnamefont {Z.}~\bibnamefont
  {Guo}}, \bibinfo {author} {\bibfnamefont {F.}~\bibnamefont {Jia}}, \bibinfo
  {author} {\bibfnamefont {L.}~\bibnamefont {Li}}, \bibinfo {author}
  {\bibfnamefont {Y.}~\bibnamefont {Ma}}, \bibinfo {author} {\bibfnamefont
  {J.~M.}\ \bibnamefont {Hutson}}, \bibinfo {author} {\bibfnamefont
  {X.}~\bibnamefont {Cui}}, \ and\ \bibinfo {author} {\bibfnamefont
  {D.}~\bibnamefont {Wang}},\ }\href {\doibase
  10.1103/PhysRevResearch.3.033247} {\bibfield  {journal} {\bibinfo  {journal}
  {Phys. Rev. Res.}\ }\textbf {\bibinfo {volume} {3}},\ \bibinfo {pages}
  {033247} (\bibinfo {year} {2021})}\BibitemShut {NoStop}%
\end{thebibliography}%

\end{document}